# Selective Encryption of VVC Encoded Video Streams for the Internet of Video Things

Amir Fotovvat and Khan A. Wahid, *Senior Member, IEEE*

*Abstract*— Visual sensors serve as a critical component of the Internet of Things (IoT). There is an ever-increasing demand for broad applications and higher resolutions of videos and cameras in smart homes and smart cities, such as in security cameras. To utilize this large volume of video data generated from networks of visual sensors for various machine vision applications, it needs to be compressed and securely transmitted over the Internet. H.266/VVC, as the new compression standard, brings the highest compression for visual data. To provide security along with high compression, a selective encryption method for hiding information of videos is presented for this new compression standard. Selective encryption methods can lower the computation overhead of the encryption while keeping the video bitstream format which is useful when the video goes into untrusted blocks such as transcoding or watermarking. Syntax elements that represent considerable information are selected for the encryption, i.e., luma Intra Prediction Modes (IPMs), Motion Vector Difference (MVD), and residual signs., then the results of the proposed method are investigated in terms of visual security and bit rate change. Our experiments show that the encrypted videos provide higher visual security compared to other similar works in previous standards, and integration of the presented encryption scheme into the VVC encoder has little impact on the bit rate efficiency (results in 2% to 3% bit rate increase).

## I. INTRODUCTION

IN the coming years, the Internet of Things (IoT) will become a key technology in the world. The purpose of IoT would be to connect a variety of end nodes and devices to the Internet in order to exchange data and to help many tasks become automated. New advances in IoT will bring many opportunities along with many challenges. Internet of Video Things (IoVT) is an essential subset of the IoT which can be defined as the internetworking of visual sensors [1]. Visual sensors generate versatile and richer data; with the advances in machine vision and artificial intelligence, cameras are becoming a favorite component in IoT systems. For example, automated retail stores such as Amazon Go are employing cameras and sensors instead of cashiers and workers for giving services to the shoppers. Other broad areas where IoVT can be applied to are home security, video surveillance, and smart city. Compared to other IoT sensors and devices, ubiquitous application of visual sensors will lead to more challenges since the storage, computation, transmission, and privacy of large volumes of videos would require significant considerations [1].

Considering the growing demand for IoVT and video streaming, the amount of video data on the Internet is rapidly increasing. In [2], It is predicted that by 2030, 13 billion cameras will be around the world from which a huge amount of data will be generated. Cisco predicts that in 2022, 82% of IP traffic will be consumed by video content [3]. Another report indicates a significant demand for video content in future smart homes [4]; Fig. 1 shows the estimated bandwidth requirement from the results of this report (applications sorted based on their predicted appearance time in future, the upper it is the earlier appearance time would be). With the upcoming video technologies and the demand for higher bandwidth, it is necessary to provide efficient solutions for privacy, storage, and transmission of video data. There are two general methods to encrypt a video prior to transmission, naive encryption and selective encryption. Naive encryption is defined where the whole video bitstream is encrypted. Using a secure encryption algorithm, this type of video encryption would completely hide any information within the video. On the contrary, selective encryption is when only some of video elements are being encrypted. Fig. 2 shows an overview of a simplified IoT environment with visual sensors. Depending on the application and different scenarios, some of the computations, such as encryption and compression, can be processed at the edge. Then, the video data can be transmitted to the cloud for other processing, such as machine vision applications and storage. So, while a video is being compressed at the edge, some of the most important elements of the encoder can be encrypted to hide the information of the video, then the encrypted video can be transmitted to the cloud servers.

Naive encryption is preferred in applications where complete confidentiality of videos is the top priority since it hides all the video information. Selective encryption can lower computation complexity and also is a suitable choice for situations where the video stream is going into other untrusted blocks (for applications such as transcoding, watermarking, and cutting) while being transmitted through the network [5]. This is due to the fact that, unlike naive encryption, videos encrypted by selective encryption approaches are decodable and can be treated like a normal video stream [6]. These features would make selective encryption a suitable choice for Digital Right Management (DRM) services as well [7]. Such SE algorithms can provide a sufficient protection level with relatively small

The Authors are with the Department of Electrical and Computer Engineering, University of Saskatchewan, Saskatoon, SK S7N 5A9, Canada (e-mail: a.fotovvat@usask.ca; khan.wahid@usask.ca).

2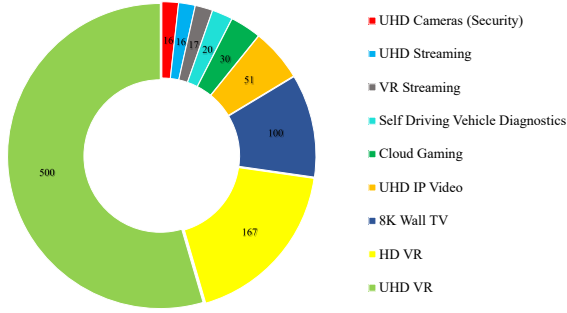

Fig. 1. Bandwidth demand for future connected homes (Mbps).

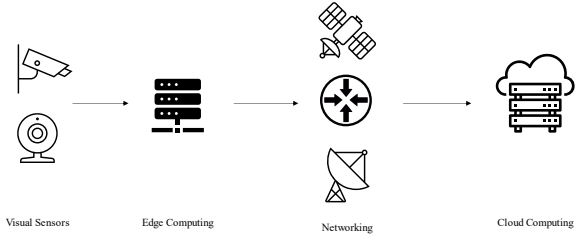

Fig. 2. IoT environment with visual sensors.

computation cost and delay. The level of output visual deterioration in SE algorithm is completely dependent to the number and type of syntax elements that are selected for the encryption. Thus, to hide video information using this methods, most important elements of the encoder that carry significant information of the video such as Motion Vector Differences (MVD), MVD signs, and Intra Prediction Modes (IPM) are usually encrypted. The encryption of syntax elements in the encoder should not cause the decoder to behave strangely, or in other words the encryption should not change the format compliancy of the video bitstream so that it can be decoded by the receiver with no problem. Thus, for the selection of syntax elements in the selective encryption methods, format compliancy of encoder and decoder is of high importance. This paper proposes a selective encryption algorithm for the H.266/VVC which is the latest video compression standard. This next generation of compression standard provides up to 50% more compression rate than the H.265/HEVC [8], thus it can play an important role to reduce the bandwidth consumption or storage requirement of video contents. In the following sections of the paper, first we will provide an overview of the related works in part one of section two, then in part two VVC coding and some of its new algorithms and improvements over HEVC will be discussed. In section three, the methodology of tests along with discussions on the presented selective encryption scheme are presented. Finally, section four covers the discussions of the experimental results.

## II. Encryption of Encoded Video Streams

### A. Overview of Related Works

Among the related literature, there are a variety of proposed methods that discuss how to securely hide video content information. The encryption process can be performed after video encoding where we encrypt some or all of the encoded bitstream, it can be integrated inside the compression algorithm (selective encryption), or the encryption can be performed by scrambling the pixels with chaos-based algorithms. Thus, the performance and characteristics of these methods essentially depend on the domain we apply the encryption to [6]. In [9], for the encryption of MPEG-1 bitstreams, it is proposed to encrypt the complete bitstream or important parts such as headers depending on the desired security level. Since this is considered naive encryption, the format compliance will not be maintained. Muhammad *et al.* [10] present a method for encryption of visual data in IoT systems. They use probabilistic algorithms to detect frames with a required level of abnormality, then pixels of these keyframes are encrypted using chaotic maps and pseudorandom number generators (PRNGs). Based on our experiments, we observed that depending on the number of encrypted frames, chaotic encryption algorithms can become quite expensive in terms of computation cost because of the processes used for the scrambling of pixels. In another paper, Preishuber *et al.* also show some other problems with chaos-based encryption algorithms [11]. Yang *et al.* [12], propose discrete sine and cosine transforms for the purpose of video encryption in the H.264/AVC. In [13], residual component, intra and inter prediction modes, and MVD components are used for the encryption in H.264/AVC standard. Wallendael *et al.*[14]

TABLE I. SOME OF RELATED WORKS FOR VIDEO ENCRYPTION

| Video Encryptin Scheme | Compression Standard | Cipher | Elements Used for Encryption | Bitrate Increase | Format Compliancy |
|---|---|---|---|---|---|
| [6] | H.264/AVC - H.265/HEVC | AES-CTR | Luma IPM, MVD sign and values, MV reference idx, Merge idx, MVP idx, Residual sign and values, SAO filter | Yes | Yes |
| [9] | MPEG-1 | DES-CBC | Headers, DCT coefficients, or whole bitstream (based on security levels) | No | No |
| [10] | N/A | PRNG | Pixels | No | Yes |
| [14] | H.265/HEVC | AES | MVD sign and values, Residual sign and values, delta QP, reference pic idx, Merge idx, MVP idx, SAO parameter | Yes | Yes |
| [15] | H.265/HEVC | AES-CTR | Luma and chroma IPM, QTC, MVD signs and values | Yes | Yes |
| [16] | H.265/HEVC | RC4 | QTC, MVD sign and values, IPM luma | Yes | Yes |
| [17] | H.265/HEVC | AES-CTR | Chroma and Luma IPM, Residual Sign and values, MVD sign and values, Merge Idx, MVP idx, Reference frame idx, SAO parameter | Yes | Yes |

investigated the encryption of syntax elements in the HEVC encoder, such as MVD values, delta QP, and residual components. They also discussed the effect of encryption on the encoder bit rate and visual scrambling of output frames. In [6], the authors presented an SE algorithm for The Context-Aware Binary Arithmetic Coding (CABAC) in the H.264/AVC and H.265/HEVC standards. They consider encryption of a wider range of elements in the aforementioned standards, including Luma IPM components. The presented experimental results of their work show the impact of the selective encryption on the encoder bit rate and visual distortion of video frames. Thiyagarajan *et al.* [15] proposed a more efficient SE algorithm for the Internet of multimedia things. They used a method to estimate the energy level of frames, then based on that energy level the algorithm decides what syntax elements of the H.265/HEVC standard be used for the encryption. Although it should be noted that the estimation of texture and motion energy levels in each frame is adding extra overhead, thus the real efficiency of the proposed system will get lower. Obviously, because their method is not encrypting all of the syntax elements in the encryption process, the reported metrics used to compare the visual distortion indicate the visual deterioration to be lower compared to other methods. Xu [16], proposes a similar encryption approach using MVD, IPM, and Quantized Transform Coefficient (QTC) elements along with a method for data embedding using QTC values in the H.265/HEVC standard. Authors of [17] tried to encrypt almost all of the important syntax elements of the H.265/HEVC standard and presented their work as a tunable approach. Also, another scrambling method is used to further distort the edges since regular SE algorithms might not completely hide edge regions; off course this added scrambling process will increase the computation complexity of the proposed method. Authors' results also indicate that the bitrate increase would be between 2% to 10% depending on whether TU coefficients are scrambled or not. Also, in Table I, a summary of some of related works on selective encryption are provided. In this paper, we are addressing the selective encryption in the H.266/VVC to see how the new improvements in this new compression algorithm affects the SE methods. Since no other work is available for SE algorithms on H.266/VVC, we will be comparing the results with previous similar works where authors were using previous compression standards.

*B. H.266/VVC Video Compression Standard*

Since 2015, there have been many efforts and activities to develop a new compression standard. It was until 2020 that the Joint Video Exploration Team (JVET) of the ITU and ISO/IEC Moving Pictures Expert Group (MPEG) announced the finalization of the H.266/VVC (otherwise known as ISO/IEC 23090-3). This new compression standard provides up to 50% more compression rate compared to the H.265/HEVC standard and supports video resolutions from 4K to 16K in addition to 360-degree videos. The official reference software of H.266/VVC standard is VVC Test Model (VTM) [18]. However, recently a faster and more optimized software named VVenC was released [19]. The new standard is in fact very similar to the HEVC since it is still a hybrid video codec (see in Fig. 3). However, H.266/VVC is powered with many new coding tools along with lots of improvements and refinements over the previously implemented technologies. In terms of performance, Laude *et al.* [20] conducted a comprehensive performance comparison of VTM, AV1, HM (HEVC software), x264, and x265 video coding software. Their results show that on average, VTM gives 5% better BD-rate performance compared to AV1. For the 4K videos, VTM outperforms AV1 with 20% gain. However, the VTM encoding time is around two to three times of AV1 depending on the tested video sequences. However, we should mention that since the H.266/VVC is announced very recently, there will be more optimizations over its features and implementations in the near future. H.266/VVC is a hybrid codec and very similar to previous standards, thus, we only discuss some of the new features and optimizations of the H.266/VVC over H.265/HEVC that will affect the process of selective encryption.

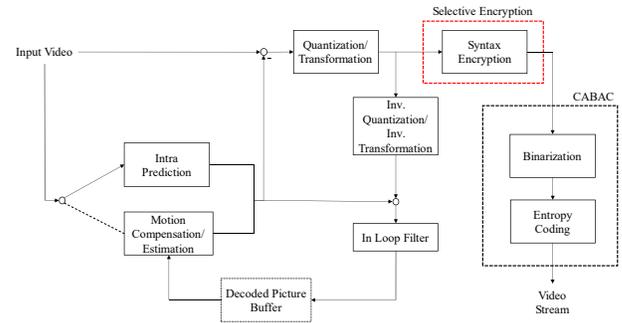

Fig. 3. Block diagram of VTM software with the selective encryption block added before CABAC.

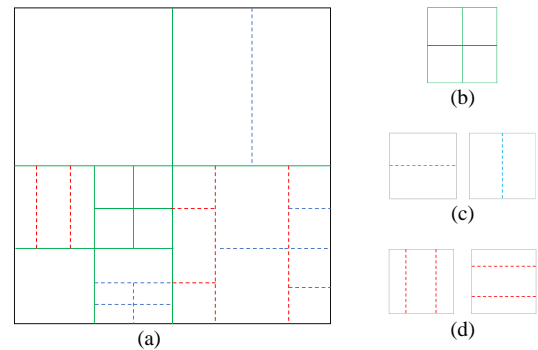

Fig. 4. (a) shows a sample block partitioning in VVC. (b) is quadtree splitting. (c) is vertical and horizontal binary mutli-type tree. (d) is vertical and horizontal ternary mutli-type tree.

During the process of compression, similar to H.265/HEVC, each frame is first split into smaller units which are named Coding Tree Units (CTUs). Size of each CTU in the H.265/HEVC standard is 64x64 pixels, but in the H.266/VVC its dimension is increased to 128x128 pixels. Then, a sequence of CTUs is grouped together in rectangular shapes to form tiles. Also, in VVC two modes of slicing are considered, rectangular slice mode and raster scan slice mode. After dividing frames into CTUs, each CTU is further divided into Coding Units (CUs). In the H.265/HEVC, during prediction and transformation CUs will also be respectively divided into smaller Prediction Units (PUs) and Transformation Units (TUs). However, concept of separating CU, TU, and PU are no longer in use (except for CUs with a larger size than the maximum transform length), in fact VVC uses CUs as the basic processing units. Moreover, for the purpose of splitting CTUs, two types of hierarchical trees are used: quadtree (QT) and multi-type tree (MTT). First, CTUs are divided with a quaternary tree then, if needed, the leaf nodes are further partitioned with multi-type tree (multi-type splitting types are shows in Fig. 4). As a result, the VVC encoder has a wide range of rectangle blocks and shapes for a CTU to split into, which would bring higher compression (as well as higher computation cost). Also, the tiles, slices, and subpictures are separated in the bitstream, which is suitable for parallel processing in encoder and decoder. This technique would also be helpful in the 360-degree videos since it allows the receiver to only decode the regions of video that the user is seeing.

For Intra prediction, there are 67 Intra modes in the H.266/VVC which make the prediction more accurate compared to H.265/HEVC standard which has 35 modes. Two of the 67 modes are planar and DC modes, the rest belong to the angular predictions. In the new coding standard, the intra prediction is similar to H.265/HEVC. Depending on the upper and left blocks of the current coding block, a list of Most Probable Modes (MPMs) with size of six, will be constructed. Then, if the current prediction mode is one of MPM elements, the encoder binarizes its index using truncated unary coding, otherwise index of the remaining 61 modes will be binarized. Because the partitioning blocks in H.266/VVC are not necessarily squares, some of prediction angles have become wider than usual 45 degrees to -135 degrees. This would allow using more reference pixels for the prediction. For Inter Prediction, VVC has several new tools. One of the interesting new features is affine motion compensation. In previous codecs like H.265/HEVC, the movement of objects are only in 2D dimensional directions. However, in real videos we barely see only planar motions and movements of objects between video frames might accompany rotation or scaling (zooming in or out). Thus, affine motion compensation is introduced to represent these more complex motions. There are a couple of other newly introduced tools for inter prediction among which we can mention Merge Mode with MVD (MMVD), Symmetric MVD Coding, Adaptive Motion Vector Resolution (AMVR), etc. For more information and details of these feature readers can refer to VTM algorithm descriptions [21].

### III. VVC SELECTIVE ENCRYPTION

#### A. Configuration, Software, and Test Video Sequences

A wide range of video sequences, from low resolution to 4K videos with different frame rates and bit depths, are selected for our tests in order to investigate results of the proposed SE algorithm (selected test videos are shown in Table II). The software for H.266/VVC is VTM [18]; currently, its latest version is 10.2. There is another software for VVC coding named VVenC [19], which is actually much faster compared to VTM. We can look at VTM as the complete software with all of the tools and options, while VVenC provides a faster implementation of H.266/VVC. However, unfortunately, the VVenC only has the encoder, and the decoder side is not available yet. Thus, for our simplicity during the tests we used the VVenC for the encoding and VTM for decoding video streams. We have used the *randomaccess_faster* configuration of the VVenC software which has *GOPsize* of 32 and *intraPeriod* of 32. In this case there will be one I frame with 31 following B frames. The base quantization parameter (QP) is also selected to vary from values 8, 24, and 32 depending on the performed tests. The *level* value which determines the tire that the encoded bitstream compiles to is chosen from Table 140 and Table 141 of [22] for each corresponding test video. Its value is dependent to video resolution (luma height and width) and frame rate of the video under test.

TABLE II. VIDEO SEQUENCES USED FOR EXPERIMENTS

| Video | Resolution | Frame Rate | Bit depth |
|---|---|---|---|
| Mobile | 352x288 | 24 | 8 |
| BasketballPass | 416x240 | 50 | 8 |
| BQMall | 832x480 | 60 | 8 |
| Johnny | 1280x720 | 60 | 8 |
| FourPeople | 1280x720 | 60 | 8 |
| BasketballDrive | 1920x1080 | 50 | 8 |
| PeopleOnStreet | 2560x1600 | 30 | 8 |
| Traffic | 2560x1600 | 30 | 8 |
| RaceNight | 3840x2160 | 50 | 10 |
| HoneyBee | 3840x2160 | 120 | 10 |

#### B. Encryption of Syntax Elements

The encryption is applied prior to the binarization of syntax elements (as shown in Fig. 3). Similar to HEVC, Context-Aware Binary Arithmetic Coding (CABAC) in VVC has two *regular* and *bypass* modes for encoding. In *bypass* mode, all symbols are considered as equiprobable for the encoding. For *Regular* mode, a probability model is determined using context of elements and the encoding is performed based on this probability model. Thus, encryption of syntax elements encoded with bypass mode would be more suitable since it does not affect the probability model, which would lower the encoding efficiency. For the encryption of elements that are encoded in *regular* mode, we should also make sure that their probability model is also changed accordingly. In the proposed method, the key syntax elements that are considered for the encryption in the H.266/VVC algorithm are luma IPM value, horizontal and vertical values/signs of motion vectors, and signs of residual values. For the encryption process, a binary stream needs to be generated using an encryption algorithm, and then the XOR operation is made between the stream and syntax elements. For this purpose, it is recommended to use the Advanced Encryption Standard (AES) in CTR mode of operation for the encryption and decryption. The x and y position values of each unit can be used for the Initialization Vector (IV) of the encryption algorithm (e.g., AES-CTR), which will maintain the security of key during the encryption of all syntax elements. Since some of the syntax elements have a specific range (e.g., luma most probable modes are from 0 to 5), the XOR operation should be performed carefully so that the final value is not outside of the available range of the corresponding syntax element. Without such considerations, because of format compliancy of the video bitstream, the decoder might face some problems. Thus, we choose a desirable number of last bits in the binary stream generated from the AES-CTR and then we do the XOR operation if its value after XOR is not beyond the range of available values. For the encryption of luma IPM modes, the encoder either encodes the index of Most Probable Mode (MPM) (predicted from the neighboring units) or it encodes the index of 61 remaining modes for the binarization (when the mode is not among the MPM modes). Encoding of the MPM uses *bypass* mode and requires five bits since it uses truncated unary coding for binarization. For the remaining IPM modes, the six bits fixed length coding is used and encoding is done in *bypass* mode. Thus, the encryption is applied to luma modes according to the selected mode for the luma intra prediction and their respective binarization. For the format compliancy, when MPM modes are being used, the encrypted syntax should not be greater than 5. Also, when the prediction mode is not among the MPM modes, the encrypted remaining IPM mode has a maximum range of 61 because six MPM modes are removed from all 67 available luma modes. Motion vectors constitute of coefficients and corresponding signs for horizontal and vertical directions. The absolute values are encoded using two flags in *regular* mode, (*abs_mvd_greater_0* and *abs_mvd_greater_1*) and *abs_mvd_minus_2* which is *bypass* coded with golmb-rice binarization. We only have chosen to encrypt the *abs_mvd_minus_2* and corresponding signs since they are encoded in *bypass* mode. Another syntax element selected for encryption is *signPattern* which represents signs of residual coefficients and is encoded in *bypass* mode.

## IV. EXPERIMENTAL RESULTS

### A. Visual Security

The encryption of selected syntax elements results in highly distorted videos which indicates that the details and information of videos are securely hidden. Fig. 5 shows the visual disturbance caused by the encryption of each syntax element. Based on these sample frames, we observe the importance of luma IPM modes in the H.266/VVC videos since their encryption causes most of deterioration in the output video (as can be seen in the column four of the Fig. 5). After luma modes, MVD modes are the next key syntax elements that their encryption is causing highest deterioration in final videos (column three Fig. 5). However, residual signs have a smaller impact on the output video as can be seen from column two of Fig. 5. The quantitative performance of the videos are compared using PSNR, SSIM [23], and VMAF [24], [25]. SSIM shows the structural similarity index which compares the processed images (the encrypted video frames for our application) and the original images (when the video is encoded with VVC but is not encrypted). To measure the performance of an encryption algorithm, the lower the SSIM the better the performance of the algorithm would be. PSNR is another metric which is not very good four our application, however because of its popularity we have employed this metric as well. Video Multimedia Assessment Fusion (VMAF) is a relatively new metric used for video quality measurement which is developed by Netflix. The VMAF returns a score between 0 to 100, and a higher score shows more resemblance between the two given original and processed videos. In table III, the performance results from the mentioned metrics are provided for selected test videos while using the encryption. We can see that the SSIM values for encrypted videos are much lower compared to the original values. The PSNR values also confirm the effect of encryption on visual deterioration of encrypted videos. In some cases, we observe mediocre PSNR scores even though the SSIM is very low and video quality is mostly disturbed. This shows the fact that PSNR might not be a very good metric here. Also, the reason that the VMAF scores for some original videos are not as high as it should be (close to 100) is that we have not used the 4K models. VMAF is in fact a metric that is trained with resolutions up to 1080p, as a result we see mediocre scores for some of original videos. However, these scores can still be used as a comparative metric between videos. The reported values in the Table III are achieved from the average of first 64 frames in the selected test videos, however In Fig. 6, we can see the SSIM, PSNR, and VMAF results for each of first 64 frames. When only encrypting the MVD elements (signs and values for horizontal and vertical directions), the I frames would not have enough visual disturbance (e.g., frame 32 in Fig. 6). However, Encryption of Luma components results in a steady highly deteriorated output for all frames. Moreover, because the residual signs are not highly important in hiding video information, only encrypting them will not yield acceptable disturbance in the encrypted videos. Because our work is the first one that looks at selective encryption in the H.266/VVC, we only compare the results with other similar methods that are used in H.265/HEVC and H.264/AVC compression standards.

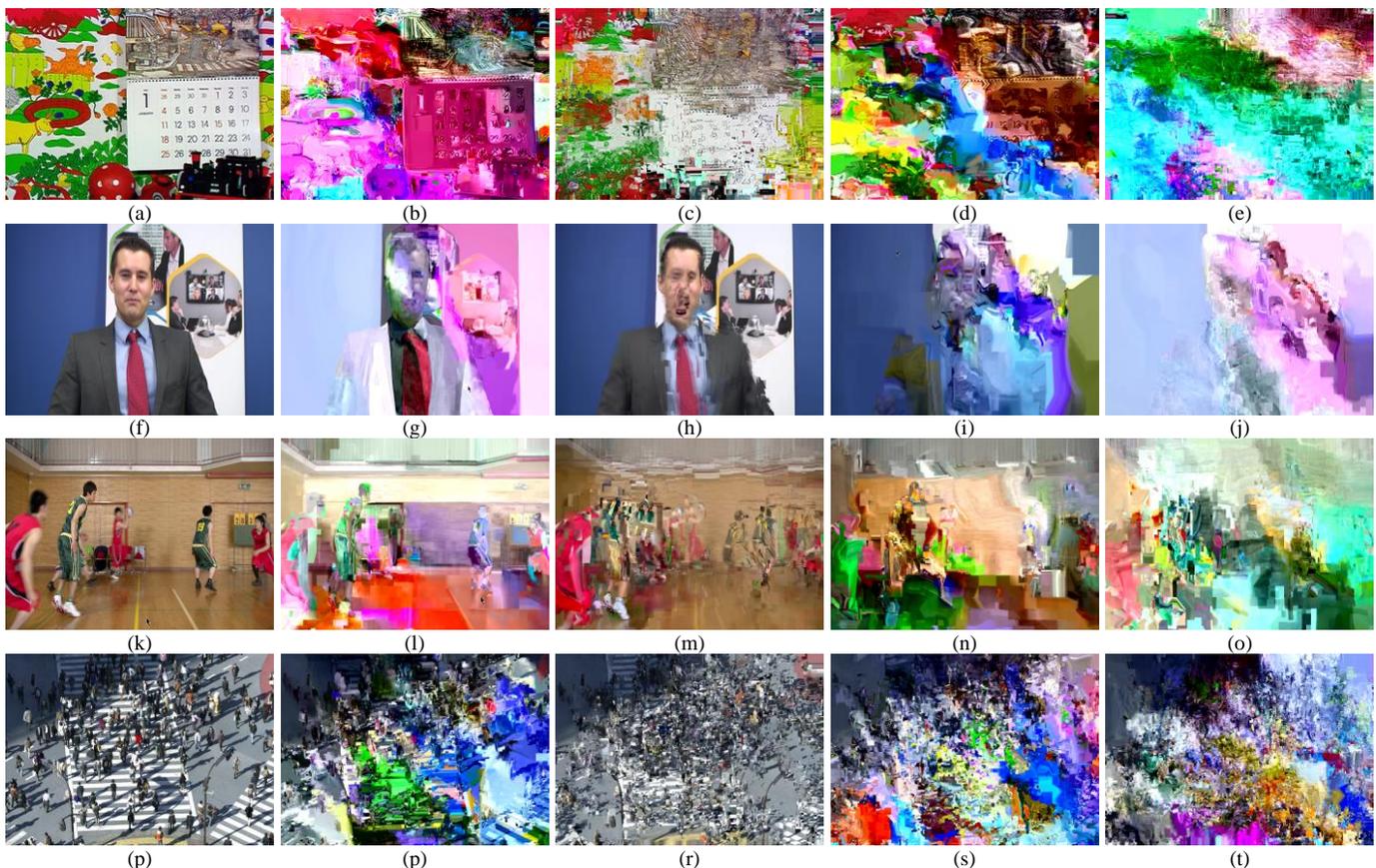

Fig. 5. Visual results of experimental tests. First column shows the original frames, second column is when chroma modes are encrypted, third column is when luma modes are encrypted, forth column is when MVD values and signs are encrypted, and fifth column is when all previous elements are encrypted together.

In the Table. IV, we see a comparison with some of other recent papers for SE algorithms in H.265/HEVC. From this comparison, we see the higher disturbance that is caused in the VVC encoded videos by presented selective encryption algorithms. In most of the cases we notice the performance of selective encryption to be higher compared with state of the art works in HEVC standard. In [17], authors select almost all of key syntax elements for the encryption in the H.265/HEVC video encoder but here we are encrypting only some of key syntax elements. If wider range of elements were considered in our scheme for the selective encryption, the quantitative results should even become better. Also, we should mention that the computation cost depends on how many syntax elements and how many units are selected for the encryption. the ideal case would be when the video disturbance is high, i.e., high security of video, while the computation cost remains low.

TABLE III – PERFORMANCE OF THE SCHEME IN TERMS OF SSIM, PSNR AND VMAF FOR EACH VIDEO SEQUENCE.

| Video Sequence | QP | Original | | | Encrypted | | |
|---|---|---|---|---|---|---|---|
| | | SSIM | PSNR | VMAF | SSIM | PSNR | VMAF |
| Mobile | 8 | 0.997 | 49.81 | 99.959 | 0.078 | 9.235 | 6.642 |
| | 24 | 0.98 | 35.78 | 99.234 | 0.075 | 8.819 | 6.190 |
| | 40 | 0.876 | 26.25 | 75.375 | 0.072 | 9.027 | 6.068 |
| BasketballPass | 8 | 0.995 | 50.52 | 99.339 | 0.362 | 13.21 | 2.877 |
| | 24 | 0.972 | 40.48 | 94.397 | 0.419 | 15.01 | 2.254 |
| | 40 | 0.834 | 30.51 | 52.556 | 0.43 | 10.17 | 2.2 |
| BQMall | 8 | 0.999 | 49.88 | 99.959 | 0.201 | 10.62 | 5.015 |
| | 24 | 0.989 | 38.47 | 98.745 | 0.216 | 10.81 | 5.457 |
| | 40 | 0.912 | 29.39 | 66.379 | 0.227 | 9.8 | 6.445 |
| Johnny | 8 | 0.998 | 50.12 | 98.248 | 0.397 | 8.596 | 0 |
| | 24 | 0.992 | 42.34 | 95.258 | 0.476 | 9.510 | 0 |
| | 40 | 0.976 | 36.47 | 79.389 | 0.557 | 11.74 | 0 |
| FourPeople | 8 | 0.999 | 50.09 | 98.339 | 0.227 | 8.75 | 0 |
| | 24 | 0.995 | 42.22 | 94.905 | 0.234 | 9.07 | 0 |
| | 40 | 0.974 | 34.71 | 75.771 | 0.266 | 9.93 | 0.005 |
| BasketballDrive | 8 | 0.999 | 50.18 | 99.958 | 0.292 | 12.95 | 1.342 |
| | 24 | 0.994 | 38.81 | 99.875 | 0.321 | 12.06 | 1.618 |
| | 40 | 0.949 | 33.08 | 65.037 | 0.340 | 11.39 | 1.826 |
| PeopleOnStreet | 8 | 0.999 | 50.25 | 99.959 | 0.085 | 9.90 | 6.658 |
| | 24 | 0.998 | 38.86 | 99.864 | 0.096 | 10.02 | 7.033 |
| | 40 | 0.977 | 29.48 | 62.668 | 0.099 | 9.79 | 6.973 |
| Traffic | 8 | 0.999 | 49.85 | 99.959 | 0.11 | 10.09 | 3.899 |
| | 24 | 0.998 | 40.65 | 95.48 | 0.126 | 9.28 | 3.815 |
| | 40 | 0.982 | 32.69 | 67.681 | 0.135 | 9.16 | 3.638 |
| RaceNight | 8 | 0.999 | 50.35 | 99.956 | 0.314 | 10.57 | 0 |
| | 24 | 0.997 | 37.37 | 96.456 | 0.355 | 7.54 | 0 |
| | 40 | 0.983 | 34.60 | 69.107 | 0.531 | 14.97 | 0 |
| HoneyBee | 8 | 0.999 | 50.26 | 97.536 | 0.212 | 10.14 | 8.888 |
| | 24 | 0.997 | 39.17 | 83.673 | 0.241 | 12.41 | 6.354 |
| | 40 | 0.991 | 37.67 | 65.464 | 0.259 | 9.72 | 6.445 |

TABLE IV. COMPARISON OF EXPERIMENTAL RESULTS WITH OTHER PROPOSED SELECTIVE ENCRYPTION ALGORITHMS.

| Video Sequence | QP | SSIM | | | PSNR | | |
|---|---|---|---|---|---|---|---|
| | | Boyadjis (for HEVC) [6] | State of the art (for HEVC) [17] – Enc | Presented Scheme (for VVC) | Boyadjis (for HEVC) [6] | State of the art (for HEVC) [17] - Enc | Presented Scheme (for VVC) |
| Mobile | 8 | 0.070 | **0.058** | 0.078 | 10.81 | 10.89 | **9.23** |
| | 24 | 0.077 | 0.076 | **0.075** | 10.64 | 10.53 | **8.82** |
| | 40 | 0.110 | 0.097 | **0.072** | 11.46 | 10.59 | **9.03** |
| BasketballPass | 8 | 0.320 | **0.260** | 0.362 | 15.21 | **14.89** | 15.21 |
| | 24 | 0.408 | **0.372** | 0.419 | 15.48 | **15.40** | 15.51 |
| | 40 | 0.457 | 0.459 | **0.43** | 16.39 | 16.94 | **14.17** |
| BQMall | 8 | 0.238 | 0.215 | **0.201** | 13.96 | 14.56 | **10.62** |
| | 24 | 0.301 | 0.282 | **0.216** | 14.32 | 14.54 | **10.81** |
| | 40 | 0.332 | 0.312 | **0.227** | 14.82 | 14.40 | **9.8** |
| Johnny | 8 | 0.468 | 0.449 | **0.397** | 13.38 | 13.87 | **8.60** |
| | 24 | 0.569 | 0.552 | **0.476** | 13.77 | 13.68 | **9.51** |
| | 40 | 0.592 | 0.580 | **0.557** | 13.41 | 13.40 | **11.746** |
| FourPeople | 8 | 0.325 | 0.295 | **0.227** | 12.76 | 13.13 | **8.75** |
| | 24 | 0.402 | 0.381 | **0.234** | 13.61 | 13.55 | **9.07** |
| | 40 | 0.420 | 0.389 | **0.266** | 13.07 | 12.83 | **9.93** |
| BasketballDrive | 8 | 0.492 | 0.496 | **0.321** | 14.65 | 15.21 | **12.955** |
| | 24 | 0.545 | 0.509 | **0.321** | 14.72 | 15.00 | **12.068** |
| | 40 | 0.582 | 0.560 | **0.340** | 15.49 | 15.26 | **11.391** |
| PeopleOnStreet | 8 | 0.250 | 0.232 | **0.085** | 12.93 | 13.12 | **9.90** |
| | 24 | 0.294 | 0.262 | **0.096** | 13.23 | 13.10 | **10.018** |
| | 40 | 0.332 | 0.312 | **0.099** | 13.09 | 13.23 | **9.79** |
| Traffic | 8 | 0.260 | 0.240 | **0.11** | 12.31 | 12.53 | **10.09** |
| | 24 | 0.348 | 0.328 | **0.126** | 12.81 | 12.70 | **9.28** |
| | 40 | 0.372 | 0.361 | **0.135** | 13.52 | 13.34 | **9.16** |

Edges of encrypted videos play an important factor for the visual security of encrypted video using selective encryption methods. If the selective encryption scheme is not good enough edges of the encrypted videos might leak some information about the objects in frames. In Fig. 7 you can see some of sample frames in both original (row one) and encrypted forms (row two). We can observe that an adversary cannot get any important information by considering the edges in the encrypted video. To validate the security of edges in the proposed method, we have considered the Edge Differential Ratio (EDR) as was employed in [5]. The EDR is defined as:

$$EDR = \frac{\sum_{m,n=1}^{M} | P(m,n) - \bar{P}(m,n)|}{\sum_{m,n=1}^{M} | P(m,n) + \bar{P}(m,n)|} \quad (1)$$

Where *M* is the number of edge pixels, *P(m,n)* is the value of edge pixel in the original video frame, and $\bar{P}(m,n)$ is the value of edge pixel in the encrypted video. In Table V, the average of EDR values for first 64 frames are listed for different QP values. Reported values confirm that the similarity of edges in encrypted and original frames are very little (low EDR indicate high edge similarity between given images). Note that this metric is completely dependent to the implemented edge detection algorithm. Also, from table V, we see that for QP=40, EDR values in original values is relatively high, however it does not indicate that the two compared videos are very different.

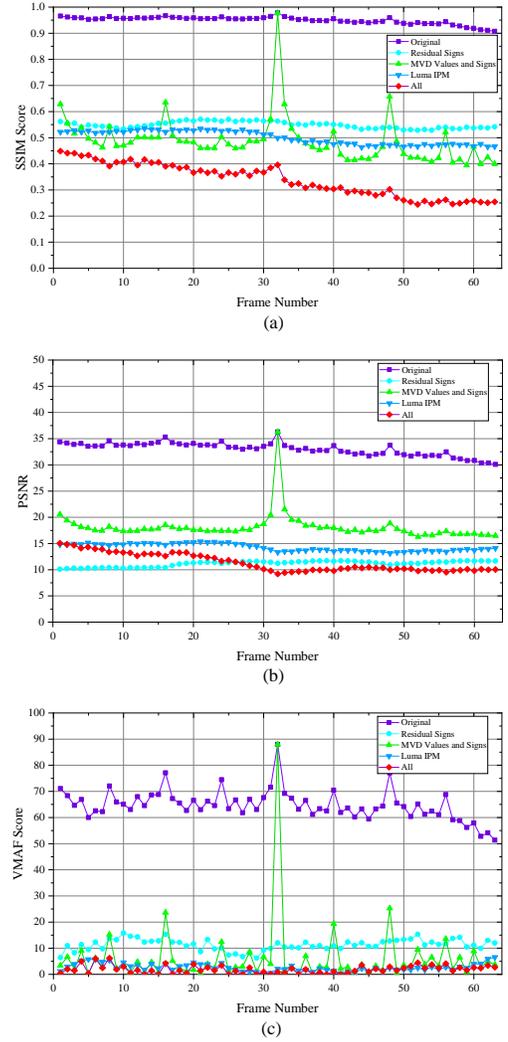

Fig. 6. Visual quality results after encrypting selected syntax elements for first 64 frames in BasketballDrive video sequence.

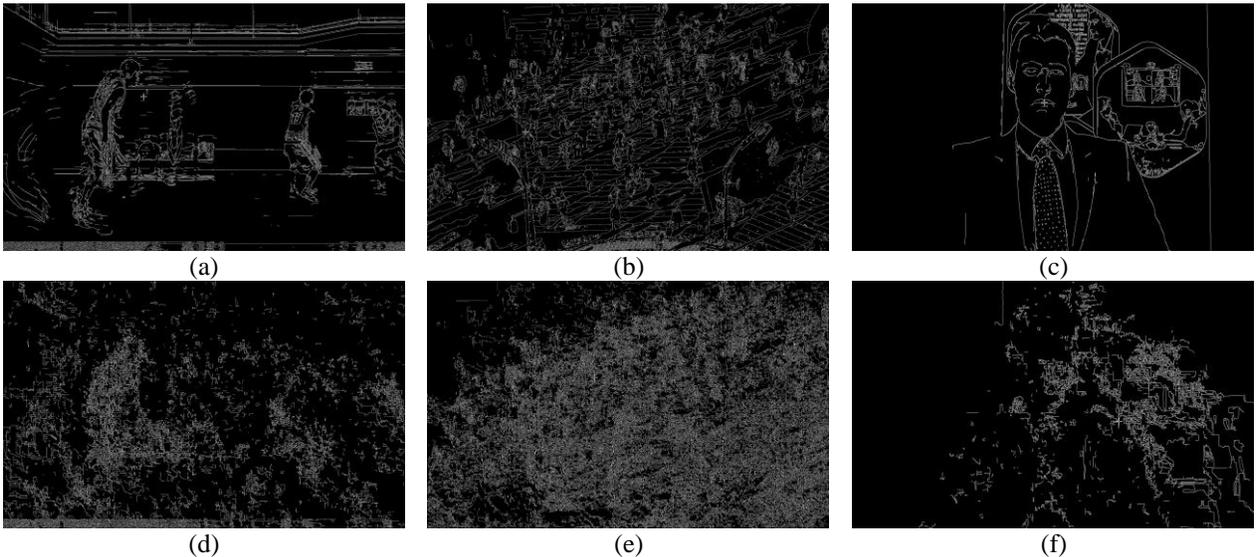

Fig. 7. Edges shows in some of video test sequences after and before performing video encryption. Images in first row are the original frames and the second row shows the corresponding encrypted frames.

This is due to the fact that while using large QP values, many of edges strength will change which results in a higher EDR value.

### B. Bit Rate Change

Because SE methods are integrated into the compression algorithms, many of the coefficients and values will be changed, resulting in a different probability model for the syntax elements. This will increase the bit rate encoding (lowering the encoding efficiency). Number of encrypted elements and selected syntax elements directly affect the bit rate change. In Table VI, the bit rate change is reported for different video bitstreams and is compared with other important related works. The results show that the bit rate increase remains around the same value as was reported for other works in H.265/HEVC, however we the presented scheme for the H.266/VVC results in more visual deterioration.

TABLE V. AVERAGE EDR SCORES OF ENCRYPTED AND ORIGINAL FRAMES

| Video Sequence | EDR | | | | | |
|---|---|---|---|---|---|---|
| | QP = 8 | | QP = 24 | | QP = 40 | |
| | Org. | Enc. | Org. | Enc. | Org. | Enc. |
| Mobile | 0.035 | 0.952 | 0.135 | 0.98 | 0.353 | 0.992 |
| BasketballPass | 0.108 | 0.91 | 0.255 | 0.928 | 0.433 | 0.968 |
| BQMall | 0.082 | 0.911 | 0.266 | 0.944 | 0.531 | 0.968 |
| Johnny | 0.239 | 0.911 | 0.357 | 0.964 | 0.449 | 0.967 |
| FourPeople | 0.194 | 0.931 | 0.35 | 0.946 | 0.535 | 0.938 |
| BasketballDrive | 0.141 | 0.781 | 0.296 | 0.985 | 0.452 | 0.986 |
| PeopleOnStreet | 0.192 | 0.946 | 0.316 | 0.98 | 0.557 | 0.991 |
| Traffic | 0.195 | 0.957 | 0.396 | 0.987 | 0.444 | 0.983 |

TABLE VI. BIT RATE CHANGE OF OUR METHOD COMPARED TO OTHER WORKS.

| Video Sequence | Bit Rate Increase | | |
|---|---|---|---|
| | Boyadjis (for HEVC) [6] | State of the art (for HEVC) [17] – Enc | Presented Scheme (for VVC) |
| Mobile | 0.0177 | 0.0244 | **0.0143** |
| BasketballPass | **0.0263** | 0.0430 | 0.0266 |
| BQMall | 0.0216 | 0.0320 | **0.0204** |
| Johnny | **0.0200** | 0.0345 | 0.0204 |
| FourPeople | 0.0246 | 0.0348 | **0.022** |
| BasketballDrive | **0.0119** | 0.0290 | 0.0179 |
| PeopleOnStreet | **0.0151** | 0.0292 | 0.0313 |
| Traffic | **0.0164** | 0.0249 | 0.0174 |

### C. Encryption Space

The available encryption space is an important factor in selective encryption algorithms since it determines the number of encrypted syntax elements. Encryption space should be large enough to make the brute force process difficult for an adversary who wants to find the keys used for the encryption. Table VII shows the encryption space used in our method compared with two other methods proposed for HEVC. It can be seen that depending on the video sequence and assigned QP value, the number of syntax elements in the presented method is 10 to 20 times larger compared to selective encryption methods in previous standards. This is probably one of the reasons that selective encryption in H.266/VCC provides higher visual security (as we discussed earlier). Moreover from Table VII, we notice a sharp change in encryption space as QP values changes. The higher QP value results in higher compression and lower output quality, which causes to have less syntax elements being encrypted (as more coefficients become zero).

TABLE VII. Comparison of Encryption Space for one I and three B frames.

| Video Sequence | QP | Selective Encryption Algorithms | | |
|---|---|---|---|---|
| | | Boyadjis (for HEVC) [6] | State of the art (for HEVC) [17] – Enc | Presented Scheme (for VVC) |
| Mobile | 8 | 408280 | 420276 | 2970442 |
| | 24 | 91249 | 98087 | 1075159 |
| | 40 | 15129 | 16234 | 285866 |
| BasketballPass | 8 | 156381 | 162952 | 1747539 |
| | 24 | 25586 | 27982 | 525422 |
| | 40 | 2694 | 3207 | 69143 |
| BQMall | 8 | 1335708 | 1369545 | 12605551 |
| | 24 | 123706 | 133542 | 1887324 |
| | 40 | 15913 | 17993 | 341932 |
| Johnny | 8 | 1529922 | 1576207 | 16440069 |
| | 24 | 67579 | 72493 | 1502727 |
| | 40 | 9478 | 10669 | 188802 |
| FourPeople | 8 | 1581943 | 1639071 | 15812535 |
| | 24 | 109661 | 115938 | 1947921 |
| | 40 | 19176 | 20552 | 371282 |
| BasketballDrive | 8 | 5747621 | 5842386 | 74431080 |
| | 24 | 248473 | 268004 | 6289156 |
| | 40 | 25455 | 30993 | 428516 |
| PeopleOnStreet | 8 | $1.05 \times 10^7$ | $1.08 \times 10^7$ | 111624820 |
| | 24 | 1144257 | 1276675 | 19361562 |
| | 40 | 136658 | 169194 | 2799887 |
| Traffic | 8 | 9217222 | 9523708 | 100424218 |
| | 24 | 782504 | 844394 | 14478533 |
| | 40 | 103101 | 110918 | 1957091 |

## V. CONCLUSION

Considering the future video technologies in smart cities, employment of high-resolution videos for streaming, security cameras, and industry applications are necessary. Hereby there is a need for works to address the security of videos. In this paper a selective encryption method for videos encoded with H.266/VVC standard is presented. Key syntax elements of the encoder such as MVD values and signs, IPM modes, and residual signs that carry important information are selected for the encryption process. Then, the integrating of presented selective encryption algorithm to the encoder is discussed both in terms of encoder performance (i.e. bitrate change), and visual security of the output video (i.e. SSIM, PSNR, VMAF). The results show that this method can be a suitable solution while maintaining security of videos for most of the applications. The importance of such selective encryption algorithms is even more noticeable when dealing with videos that require large bandwidth like 4K/8K and 360 degree videos. Future works will be about the applicability of other syntax elements in the H.266/VVC for the selective encryption and to search for more efficient methods where encryption of syntax elements is only applied to a portion of CU units to reduce the computation cost while yielding the highest visual security.